\documentclass[useAMS,usenatbib]{mn2e}
\usepackage{epstopdf}
\usepackage{amssymb,amsmath}
\usepackage{graphicx,color}
\newcommand{\aap}{A \& A}
\newcommand{\mnras}{MNRAS}
\newcommand{\y}{}
\title[Radio Emission from Discs]{Testing protoplanetary disc dispersal with radio emission}
\author[Owen, Scaife \& Ercolano]{James E. Owen$^{1}$\thanks{E-mail: jowen@cita.utoronto.ca}, Anna M.M. Scaife$^{2}$ and Barbara Ercolano$^{3,4}$\\
$^{1}$Canadian Institute for Theoretical Astrophysics, 60 St. George Street, Toronto, M5S 3H8, Canada.\\
$^{2}$School of Physics \& Astronomy, University of Southampton, Highfield, Southampton SO17 1BJ, England.\\
$^{3}$Excellence Cluster Universe, Boltzmannstr. 2, D-85748 Garching, Germany\\
$^{4}$Universit\"ats-Sternwarte M\"unchen, Scheinerstrasse 1, D-81679 M\"unchen, Germany}
\begin{document}

\pagerange{\pageref{firstpage}--\pageref{lastpage}} \pubyear{2002}

\maketitle

\label{firstpage}

\begin{abstract}
We consider  continuum  free-free radio emission from the upper atmosphere of protoplanetary discs as a probe of the ionized luminosity impinging upon the disc. Making use of previously computed hydrodynamic models of disc photoevaporation within the framework of EUV and X-ray irradiation, we use radiative transfer post-processing techniques to predict the expected free-free emission from protoplanetary discs. In general, the free-free luminosity scales roughly linearly with ionizing luminosity in both EUV and X-ray driven scenarios, where the emission dominates over the dust tail of the disc and is partial optically thin at cm wavelengths. 
We perform a test observation of GM Aur at 14-18 Ghz and detect an excess of radio emission above the dust tail to a very high level of confidence. The observed flux density and spectral index are consistent with free-free emission from the ionized disc in either the EUV or X-ray driven scenario. Finally, we suggest a possible route to testing  the EUV and X-ray driven dispersal model of protoplanetary discs, by combining observed free-free flux densities with measurements of mass-accretion rates. On the point of disc dispersal one would expect to find a $\dot{M}_*^2$ scaling with free-free flux in the case of EUV driven disc dispersal or a $\dot{M}_*$ scaling in the case of X-ray driven disc dispersal.     
\end{abstract}

\begin{keywords}
planetary systems: protoplanetary
discs - stars: pre-main-sequence.
\end{keywords}

\section{Introduction}
Planet formation and the late stages of star formation are intimately related through the evolution of the protoplanetary disc. Understanding the evolution of the protoplanetary disc is key to understanding the final stages of star formation, as well as characterising the environment in which planets form and migrate. In particular, the disc dispersal mechanism and the time-scale on which it operates sets the time-scale in which gas planets must form, and dramatically affects the environment where terrestrial plants form and evolve. 

Observations of protoplanetary discs over the last decade have shown that at birth most stars are surrounded by gas and dust rich optically thick protoplanetary discs. By an age of 10 Myr almost all young stars are disc-less, with the median disc lifetime being $\sim 3$ Myr (Haisch et al. 2001; Hernandez et al. 2007; Mamajek 2009). Furthermore, the distribution of young stars in Near-IR (NIR) colours indicates that the transition from disc bearing to disc-less occurs rapidly, on a time-scale of $\sim10\%$ of the disc's lifetime (e.g. Kenyon \& Hartmann 1995; Luhman 2010; Ercolano et al. 2011, Koepferl et al. 2013) and it is roughly constant across spectral types for stars smaller than approximately one solar mass (Ercolano et al. 2011b). 

A small sub-set of protoplanetary discs  show a lack of opacity at NIR wavelengths, but emission similar to what would be expected from a primordial disc at Mid-IR wavelengths (Strom et al. 1989; Skrutskie et al. 1990). These `transition' discs have typically been interpreted as discs with cleared inner holes in the dust component (Calvet et al. 2002,2005; Brown et al. 2009; Kim et al. 2009; Espaillat et al. 2010; Merin et al. 2010; Andrews et al. 2011). However, it is unclear whether these discs also possess holes in their gas component as most are found to be still accreting (e.g. Espaillat et al. 2010, Andrews et al. 2011, Owen \& Clarke 2012). The discovery of `transition' discs have motivated the development of dispersal models able to clear the disc from the inside out, including photoevaporation (Clarke et al. 2001), grain growth (Dullemond \& Dominik 2005; Birnstiel, Andrews \& Ercolano 2012), giant planet formation (Armitage \& Hansen 1999) or a combination of photoevaporation and giant planet formation (Rosotti et al. 2013). A large fraction of `transition' discs have inner hole sizes and accretion rates that are consistent with clearing by photoevaporation (Alexander \& Armitage 2009; Owen et al. 2011b,2012; Owen \& Clarke 2012), while others are perhaps more consistent with truncation by an embedded planet (Espaillat et al. 2010; Andrews et al. 2011; Clarke \& Owen 2013). 

The theory of disc photoevpoartion has progressed greatly since the early models. Hollenbach et al. (1994) considered semi-analytic solutions of a pure EUV heated disc, which were updated to full hydrodynamic models by Font et al. (2004) and Alexander et al. (2006). Further calculations have indicated that X-rays (e.g. Ercolano et al. 2008; Ercolano et al. 2009; Owen et al. 2010) and FUV (e.g. Gorti \& Hollenbach 2009; Gorti et al. 2009) have a stronger influence on the disc mass loss rates than the EUV. Furthermore, Owen et al. (2012) argued that X-rays are the dominant driving process for low mass protoplanetary discs, with mass-loss rates  in the range 10$^{-10}$-10$^{-8}$ M$_\odot$ yr$^{-1}$ (Owen et al. 2011b) which are set by the star's X-ray luminosity.   

While the theory of photevaporation has improved significantly since the early models, observations of disc photoevaporation has as yet been unable to directly compare the different mass-loss models. Currently the 12.8 $\mu$m NeII line provides the best direct probe of disc photoevaporation with a clear detection of a photoevaporative flow from several objects found by Pascucci et al. (2009). However, such observations are consistent with being a fully ionized EUV flow (Alexander 2008), or a partially ionized X-ray flow (Ercolano \& Owen, 2010) as the the NeII line probes electron densities rather than mass-loss rates (Pascucci et al. 2011). Further clues are provided by the blue-shifted low-velocity component of the 6300\AA~ OI line detected in many disc bearing systems (Hartigan et al. 1995) which is consistent with production in a quasi-neutral X-ray wind (Ercolano \& Owen 2010) and not from a fully-ionised EUV wind (Font et al. 2010). However, there are other non-thermal mechanisms that can produce OI emission (Gorti et al. 2011) and until resolved observation become available, this line cannot be used as a clean probe of the photoevaporation models. Comparisons of the 15.6 $\mu$m NeIII  to the NeII luminosity may indicate the origin of the ionizing radiation (Espaillat et al. 2013), but NeIII line emission is unfortunately difficult to measure due to its very low luminosity. While comparisons of various line emission diagnostics from a large sample stars performed by 	
Szul\'agyi et al. (2012) points to the soft X-rays as the dominant heating source for the surface layers of discs (consistent with the conclusions of Espaillat et al. 2013), a direct comparisons with photoevaporation models have not been conclusively able to identify the dominant driving source. 

Recently, Pascucci et al. (2012) proposed that radio emission may provide a clean test of photoevaporation models. In particular, Pascucci et al. (2012) suggested that continuum free-free emission at $\sim$ cm wavelengths could be detectable with current radio telescopes above the dust tail of the disc, and could provide a method to test the various photoevaporation models. Continuum emission is not affected by variations in elemental abundances or difficulties in resolving line profiles that are hampering current efforts to probe photoevaporation using spectroscopy.  Indeed in the case of external evaporation of discs by nearby OB stars (e.g. Johnstone et al. 1998; Richling et al. 2000; Adams et al. 2004), radio-emission has already been used to test the various evaporation models. These radio measurements have further been used to estimate mass-loss rates (e.g. Mucke et al. 2002), since the flow structures can be resolved at other wavelengths (something not yet possible for internal photoevaporation). {\y Furthermore, Pascucci et al. (2012) used archival data of TW Hydra to demonstrate a possible excess at radio wavelengths, indicative of emission by a photoevaporative wind. Without a spectral index measurement
they were however unable to confirm such an origin.} Radio-emission from young stars has gained interest recently, in particular whether the G\"udel \& Benz relation (G\"udel \& Benz 1993) which relates X-ray and radio emission from active stars can be extended to the early stages of star formation (Feigelson et al. 1994; Guenther et al. 2000; Forbrich et al. 2007, 2011; Osten \& Wolk 2009; Forbrich \& Wolk 2013). The origin of the G\"uedel \& Benz relation it typically thought to result from spatially colocated thermal X-ray emission and non-thermal radio emission from the stellar corona. Currently observed young stars with a surrounding envelope and disc are systematically and considerably more luminous in the radio than is predicted from the G\"uedel Benz relation (e.g. Forbrich et al. 2011; Forbrich \& Wolk 2013). Suggesting that the radio emission at these early stages is perhaps not dominatned by non-thermal emission from a stellar corona. In particular the radio emission maybe thermal in nature (e.g. Scaife et al. 2011), suggesting a different emission mechanism/region for the X-ray and Radio.

In this paper, we explore the expected properties of the free-free radio emission from protoplanetary discs in the context of an EUV driven and an X-ray driven photoevaporation model. In Section~2 we derive the expected signature of free-free emission from photoevaporation models.
 In Section~3 we use hydrodynamic calculations of EUV \& X-ray driven photoevaporation to calculate the expected luminosity, spectral index and spatial distribution of the free-free emission. In section 4 we apply our formalism to a test case and present a new free-free detection from GM Aur.  In Section~5 we discuss our findings and present a systematic method in which a survey of radio emission from discs could distinguish between the various photoevaporation models. A short summary is provided in Section~6.

\section{Scaling laws and expected properties}\label{sec:theory}

Our discussion follows on from the calculations performed by Pascucci et al. (2012) who suggested that thermal free-free emission from the heated surface layers of a protoplanetary disc dominates emission at radio frequencies. In particular, they argued that the free-free luminosity scales linearly with ionizing luminosity (EUV or X-rays) and has detectable values of $\sim 100$ $\mu$Jy from discs in local star forming regions at cm wavelengths. We extend their discussion here, which we then use as the basis for interpreting the detailed numerical calculations performed in Section~\ref{sec:numerical}. 

Following Pascucci et al. (2012) the volume free-free emissivity is given by:
\begin{equation}
\epsilon_\nu=6.8\times10^{-9}\;  g_{ff} n_e^2T_e^{-1/2}\exp(-h\nu/k_B T_e) \;{\rm \mu Jy\, cm^{-1}}\label{eqn:emis}
\end{equation}
where $g_{ff}$ is the Gaunt factor, $n_e$ is the electron density, and $T_e$ is the electron temperature. The total - unattenuated - luminosity density is then found by integrating over the entire volume of the flow:
\begin{equation}
L_\nu=\int\epsilon_\nu\,{\rm d^3 r}\label{eqn:lum_int}
\end{equation}
It is useful here to introduce the gravitational radius, $R_g$, as a basic length scale for photoevaporation, given by the distance from the star at which gas at a given temperature becomes unbound from the star due to thermal motions (Hollenbach et al. 1994) and is defined as:
\begin{equation}
R_g=\frac{GM_*}{c_s^2}=8.9 {\rm ~AU}\left(\frac{M_*}{1{\rm ~M}_\odot}\right)\left(\frac{c_s}{10{\rm ~km~s}^{-1}}\right)^{-2}
\end{equation}
\noindent where $M_*$ is the mass of the star and $c_s$ is the sound speed of the gas. 
It has been shown that both the EUV (Hollenbach et al. 1994) and X-ray (Owen et al. 2012) flow properties scale with this quantity (i.e. $\rho/\rho_0=N(r/R_g)$, $T/T_0=t(r/R_g)$ etc.). The EUV case results in an  isothermal flow with a temperature of $\sim10^4$K and $R_g$ is strictly a constant for the entire disc/wind system. In the X-ray case, where the gas temperature ranges from a few 1000 to $\sim 10^4$ K the meaning of $R_g$ is less clear for a single disc. Owen et al. (2012) argue that it is best to think of $R_g$ as a mass-scaled radius where the sound-speed is a scaling constant fixed for the entire disc/wind system, which we set to 10~km s$^{-1}$ for consistency with the EUV model.  

\subsection{EUV heated disc}
In the case of an EUV heated disc, where the gas is isothermal, the electron density in the EUV heated region scales as (Hollenbach et al. 1994):
\begin{equation}
n_e(r/R_g)=\Phi_*^{1/2}R_g^{-3/2} N(r/R_g)
\end{equation}
where $N(r/R_g)$ is independent of stellar mass or EUV luminosity. Combining this with Equation~\ref{eqn:lum_int} we find:
\begin{equation}
L^{\rm EUV}_\nu\propto\Phi_*\int N^2(r/R_g)\,{\rm d}^3\left(r/R_g\right) \label{eqn:euv_scale}
\end{equation}
Thus we see that free-free emission from a EUV heated disc only has an explicit linear dependence on the EUV luminosity, and no explicit dependence on stellar mass, identical to the scaling found by Pascucci et al. (2012). We can compare this to the scaling of the mass-loss rate which is given by (Hollenbach et al. 1994; Font et al. 2004):
\begin{equation}
\dot{M}^{\rm EUV}=1.3\times10^{-10}\; {\rm M}_\odot\, {\rm yr}^{-1}\left(\frac{\Phi_*}{10^{41}\,{\rm s}^{-1}}\right)^{1/2}\left(\frac{M_*}{1{\, \rm M_\odot}}\right)^{1/2}
\end{equation}
So in the EUV case $L^{\rm EUV}_\nu\propto \dot{M}^2$. 

\subsection{X-ray heated disc}
As we shall discuss in more detail in Section~\ref{sec:numerical}, it is convenient to treat the bound and unbound sections of the X-ray heated disc separately as they have a different density and temperature dependence with stellar mass and X-ray luminosity. Owen et al. (2012) showed that the temperature structure of the flow is invariant in terms of a mass-scaled radius, namely $T=T_0\,t(r/R_g)$ and that density scales as:
\begin{equation}
n_H(r/R_g)\propto L_XR_g^{-2}N(r/R_g)\label{eqn:den_xray},
\end{equation}
where, as in the EUV case, $N(r/R_g)$ and $t(r/R_g)$ are independent of stellar mass and X-ray luminosity. Equation~\ref{eqn:lum_int} can then be re-written as:
\begin{equation}\label{eqn:x-ray_scale}
L_\nu^{X, {\rm flow}}\propto L_X^2M_*^{-1}\int X_e^2( r/R_g)\frac{N( r/R_g)^2}{\sqrt{t(r/R_g)}}\,{\rm d}^3(r/R_g)
\end{equation}
where $X_e$ is the electron fraction. Given the electron fraction is set by the ionization parameter ($\xi=L_X/nr^2$), which  is invariant in terms of $r/R_g$ (see Appendix A of Owen et al. 2012), the electron fraction is also invariant under the same mass-scaled radius. So the  integral in Equation~\ref{eqn:x-ray_scale} is independent of stellar mass and X-ray luminosity. This means the free-free emission from the flow scales as the X-ray luminosity squared and inversely with stellar mass. This is different to the derived scaling from Pascucci et al. (2012) who based their argument on ionization balance in a fixed isothermal slab and arrived at a linear scaling with X-ray luminosity. This difference results from neglecting that the density in this underlying slab would also scale linearly with X-ray luminosity as shown in Equation~\ref{eqn:den_xray}. 

In the disc's bound atmosphere, inside roughly $\sim 1$ AU the temperature is roughly isothermal (at 10$^{4}$ K) and hence the density structure is independent of X-ray luminosity. Therefore, in such a region we can balance ionizations with recombinations, in the case that free electrons from Hydrogen atoms are dominant (as will be the case at the ionization fractions we are considering $>0.01$), then such a balance is similar to that of EUV ionization-recombination balance and one finds a similar expresion to Equation~\ref{eqn:euv_scale}:
\begin{equation}
L_\nu^{X, {\rm bound}}\propto L_X \int N^2(r/R_g)\,{\rm d}^3\left(r/R_g\right) 
\end{equation}
This scenario correspond to the slab calculation performed by Pascucci et al. (2012). 

To summarise in the case that the free-free emission is dominated by the wind one finds an $L_X^2$ scaling for the luminosity, and in the case the inner bound atmosphere dominates the emission one recovers the linear scaling derived by Pascucci et al. (2012). These free-free scalings can be compared to the photoevaporation rate from the X-ray wind given by (Owen et al. 2012):
\begin{equation}
\dot{M}_X=7.9\times10^{-9}\; {\rm M}_\odot\, {\rm yr}^{-1}\left(\frac{L_X}{10^{30}\,{\rm erg s}^{-1}}\right)
\end{equation}
Thus we note even in the case where the scalings of free-free emission with input energy are the same, they will scale very differently with mass-loss, a feature that we will use later in Section~\ref{sec:discuss}. 

{\y We add that in the case that the disc experiences both strong X-ray and EUV irradiation the EUV model implicitly assumes that the X-rays will only heat and ionize the dense layers of the disc to low levels. This will results in a very low volume emission measure for this X-ray ionized region and no contribution to the free-free emission, from the X-rays in this case. In the case of an X-ray dominated driven wind with EUV irradiation, Owen et al. (2012) demonstrated that the X-ray wind is itself optically thick to the EUV irradiation and the EUV ionized portion is confined to very small radius. This results in an EUV ionized region that will have a very low volume emission measure when compared to the X-ray heated disc, due to its small size.}

\section{Numerical Calculations of free-free emission}\label{sec:numerical}
In order to investigate the radio emission from the photoevaporative wind  we use previously computed hydrodynamic calculations of  EUV and X-ray driven flows. We then post-process these density, temperature and velocity structures using radiative transfer calculations to compute the properties of free-free emission from both EUV and X-ray heated discs, using a method similar to that employed by Ercolano \& Owen (2010) to compute line emission from disc winds. The free-free emissivity is calculated using Equation~\ref{eqn:emis} and then the opacity is computed as:

\begin{equation}
\kappa_\nu=\frac{\epsilon_\nu c^2}{2k_b T_e\nu^2}
\end{equation}
assuming thermal equilibrium in the Rayleigh-Jeans limit, where $c$ is the speed of light. At the long $>$ cm wavelengths we are primarily interested in the opacity from dust in the disc/wind can be neglected.  The hydrodynamic calculations have been computed on 2D spherical polar grids, which cover a radial range from $\sim 0.01R_g$ to $\sim 10R_g$. Thus in-order to perform radiative transfer ray-tracing calculations we use symmetry about the rotation axis and mid-plane to interpolate the 2D hydrodynamic grids on to 3D logarithmic spaced Cartesian grids with size $N_x \times N_y\times N_z$ of $240\times240\times3000$ which provide  sufficiently high resolution in the inner regions to resolve the disc, flow and emission structures. In order to calculate the optical depth to the observer, we use the following method: for each cell in the 3D Cartesian grid we cast a ray through the grid at the required inclination. We then calculate the optical depth to the observer by taking the opacity to be constant throughout an intercepted cell and summing up the individual optical depths through all the intercepted cells. More specifically the optical depth from a given cell to the observer is found as:
\begin{equation}
\tau_\nu^{ijk}=\sum_{\substack{\rm intercepted \\ \rm cells}}\!\!\!\!\!\kappa_\nu\ell
\end{equation}
where $\ell$ is the path length of the ray through an intercepted cell, and the sum takes place over all cells intercepted by a given ray. Then the total luminosity density can be computed as:
\begin{equation}
L_\nu=\sum_{i,j,k}\epsilon^{ijk}_\nu\exp(-\tau_\nu^{ijk})V^{ijk}
\end{equation}
where $V^{ijk}$ is the volume of the cell.   

\subsection{EUV heated disc}
\begin{figure}
\centering
\includegraphics[width=\columnwidth]{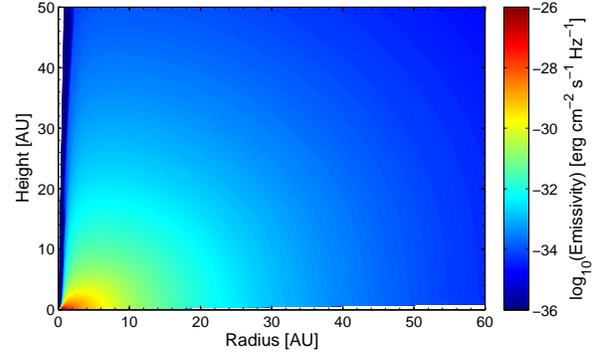}
\caption{The emissivity at 15 GHz of an EUV driven photoevaporative wind, with a luminosity of $\Phi_*=10^{41}$ s$^{-1}$ around a 0.7 M$_\odot$ star.}\label{fig:euv_emis_map}
\end{figure}

For our EUV heated disc we make use of the numerical hydrodynamic calculations of Owen et al. (2011a), which are similar to those of Font et al. (2004) and Alexander (2008). We defer to Owen et al. (2011a) for a detailed description of these calculations.  Given that EUV ionization results in a fully ionized hydrogen gas at a temperature which thermostats to $\sim 10^4$K (Hollenbach et al. 1994), we set the ionization fraction to unity and the gas temperature to a constant $10^4$K to compute the free-free emissivity. Figure~\ref{fig:euv_emis_map} shows the spatial distribution of the emissivity produced in the EUV wind: it peaks at small radius and smoothly transitions to a spherical distribution at large radius, and effectively traces the density structure of the EUV heated wind. 

\begin{figure}
\centering
\includegraphics[width=\columnwidth]{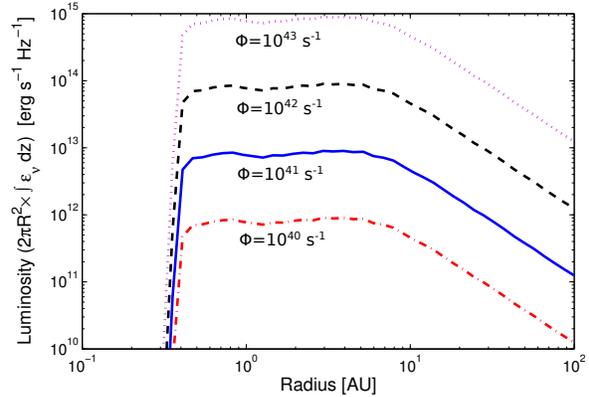}
\caption{Radial luminosity profiles at 15 GHz for the EUV driven photoevaporative wind around a 0.7 M$_\odot$, shown at several different EUV luminosities.}\label{fig:euv_emis_rad}
\end{figure}

In Figure~\ref{fig:euv_emis_rad} we show the radial distribution of the free-free luminosity for a $0.7$ M$_\odot$ star with and EUV ionising luminosity of $10^{40}$ (dot-dashed line), $10^{41}$ (solid line), $10^{42}$ s$^{-1}$ (dashed line) and $10^{43}$ s$^{-1}$ (dotted line). The lines represent the vertical integral of  $\varepsilon_\nu$ multiplied by $2\pi R^2$. The plot shows that the emission is mainly dominated by the photoevaporative wind (outside $>$ 1 AU), although a non-negligible fraction does come from the hydrostatic bound region. We also note a smooth transition in the from the bound region to the wind and a slow drop of in the luminosity at large radius. Furthermore, we see that as expected the free-free luminosity scales linearly with EUV luminosity as described in Section~\ref{sec:theory} and in Pascucci et al. (2012). 

\subsection{X-ray heated disc}
For the X-ray wind we make use of the hydrodynamic calculations performed by Owen et al. (2010,2011b,2012) and we defer to these works for a detailed discussion of the numerical hydrodynamics. We consider three X-ray luminosities of $2\times10^{29}$, $2\times10^{30}$ and $2\times10^{31}$ erg s$^{-1}$ around two stellar masses of 0.1 and 0.7 M$_\odot$. The numerical method for all the radiation-hydrodynamic calculations are described in detail in Owen et al. (2010), which considers the `standard' case with $M_*$=0.7  M$_\odot$ and $L_X=2\times10^{30}$ erg s$^{-1}$; the simulations for ten times higher and lower X-ray luminosity are described in Owen et al. (2011b) and the simulation for the lower mass star case (0.1 M$_\odot$) are described in Owen et al. (2012). 

\begin{figure}
\centering
\includegraphics[width=\columnwidth]{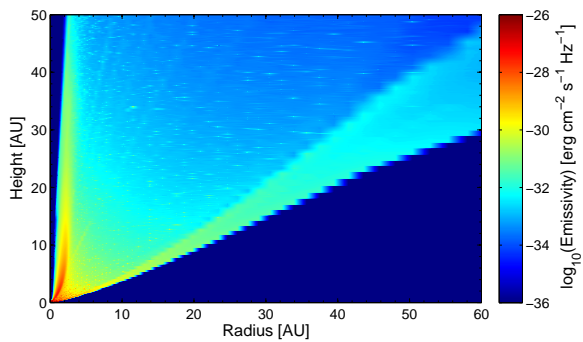}
\caption{The emissivity at 15 GHz of an X-ray driven photoevaporative wind, with an X-ray luminosity of $L_X=2\times10^{30}$ erg s$^{-1}$ around a 0.7 M$_\odot$ star.}\label{fig:xray_emis_map}
\end{figure}

Unlike the EUV case, we cannot make a priori assumptions about the electron fraction and electron temperature. The physical properties of the gas must be determined by detailed photoionization calculations, which we perform by post-processing the hydrodynamic simulations using the {\sc mocassin} radiative transfer code (Ercolano et al. 2003,2005,2008). The electron fractions and electron temperatures obtained from {\sc mocassin} are then used to calculate the free-free emissivities using Equation~\ref{eqn:emis}. The emissivity map is shown in Figure~\ref{fig:xray_emis_map} for the `standard' model. Clearly this is qualitatively different from the EUV wind case which has a smooth distribution. The emission from the wind is dominated by a narrow vertical band at close radius and it effectively traces out the absorption region of the very soft $\lesssim 0.1$ KeV X-ray photons that generate a high electron fraction and temperature in this inner region.

\begin{figure}
\centering
\includegraphics[width=\columnwidth]{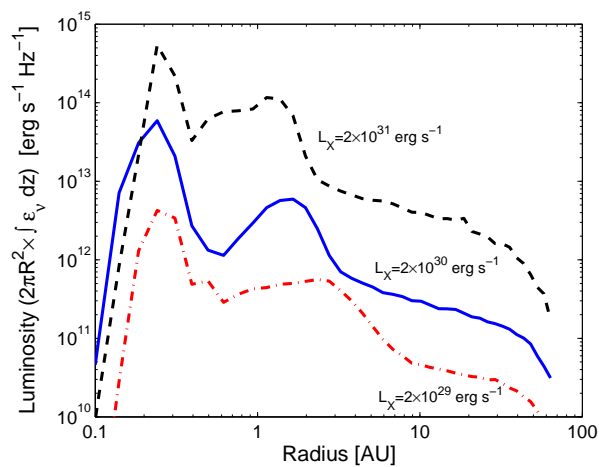}
\caption{Radial luminosity profiles at 15 GHz for the X-ray driven photoevaporative wind around a 0.7 M$_\odot$, shown for several X-ray luminosities.}\label{fig:xray_emis_rad}
\end{figure}

Furthermore, the radial distribution of the free-free emissivity is  shown in Figure~\ref{fig:xray_emis_rad} for a 0.7~M$\odot$ star with X-ray luminosity of  $2\times10^{29}$ (dash-dot line), $2\times10^{30}$ (solid line) and $2\times10^{31}$ erg s$^{-1}$ (dashed line).  Unlike the smooth distribution obtained for the EUV case, the radial luminosity profile in the X-ray case is actually double peaked. The first peak at small radius ($\lesssim 1$ AU) dominates and is produced not by the wind, but by the hot bound X-ray heated atmosphere that is unable to escape the stars potential. The second peak is at slightly larger radius ($1-10$ AU) and is produced in the wind and is only at the highest X-ray luminosities it becomes comparable to the first peak.  As expected from our discussion in Section~\ref{sec:theory}, we see that the luminosity in the bound region of the X-ray heated atmosphere scales roughly linearly with X-ray luminosity and the luminosity in the wind scales roughly as $L_X^2$, meaning that the wind begins to dominate at high X-ray luminosity. However, we note that for a total luminosity scaling roughly as $L_X^2$ one would require X-ray luminosities $\gtrsim 10^{32}$ erg s$^{-1}$ which are not observed in young solar-type stars (e.g. Guedel et al. 2007, Preibisch et al. 2005). Therefore, in general we expect the free-free luminosity to be dominated by the bound inner region and  to scale roughly linearly with X-ray luminosity.

\subsection{Spectral properties}
\begin{figure*}
\centering
\includegraphics[width=\textwidth]{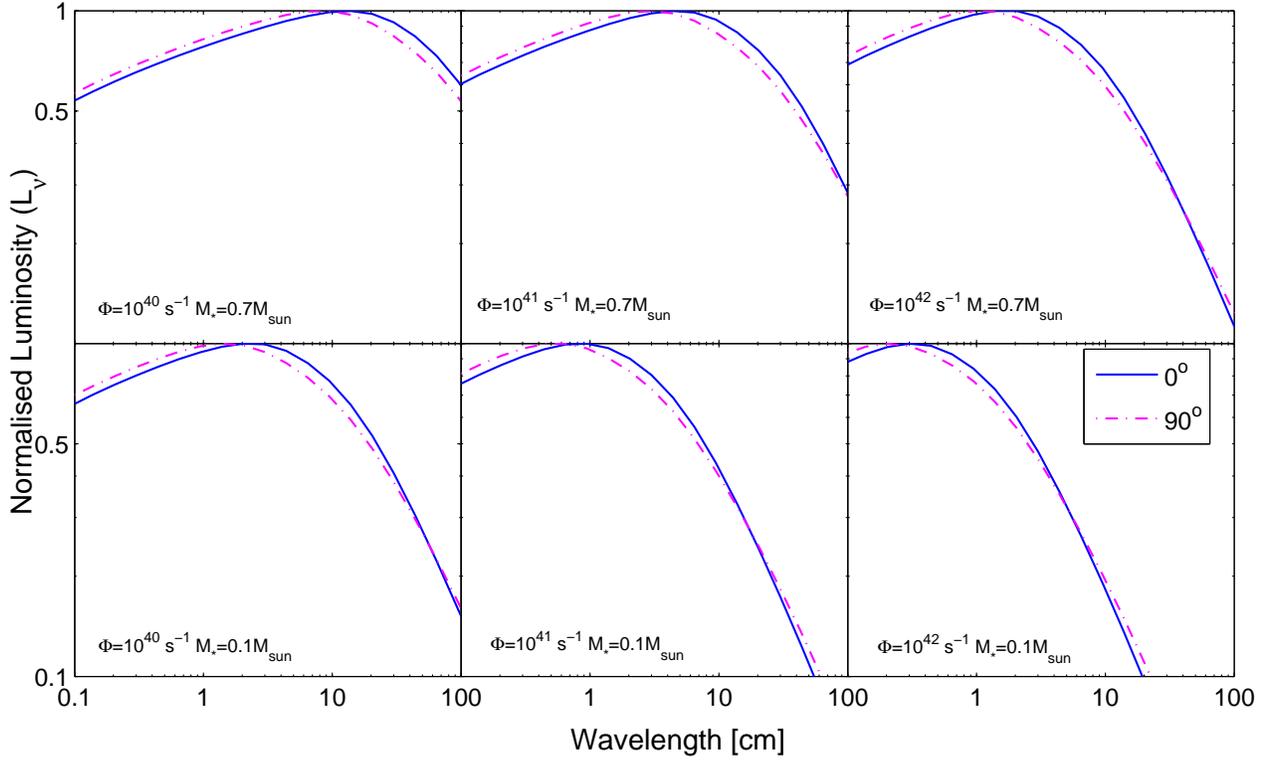}
\caption{Free-Free radio spectrum for the EUV heated wind at inclinations of 0$^o$ and 90$^o$. The top/bottom panels show the case of a 0.7/0.1 M$_\odot$ star, for EUV luminosities of $10^{40}$ s$^{-1}$ (left), $10^{41}$ s$^{-1}$ (middle) and $10^{42}$ s$^{-1}$ (right).}\label{fig:euv_spec}
\end{figure*}

\begin{figure*}
\centering
\includegraphics[width=\textwidth]{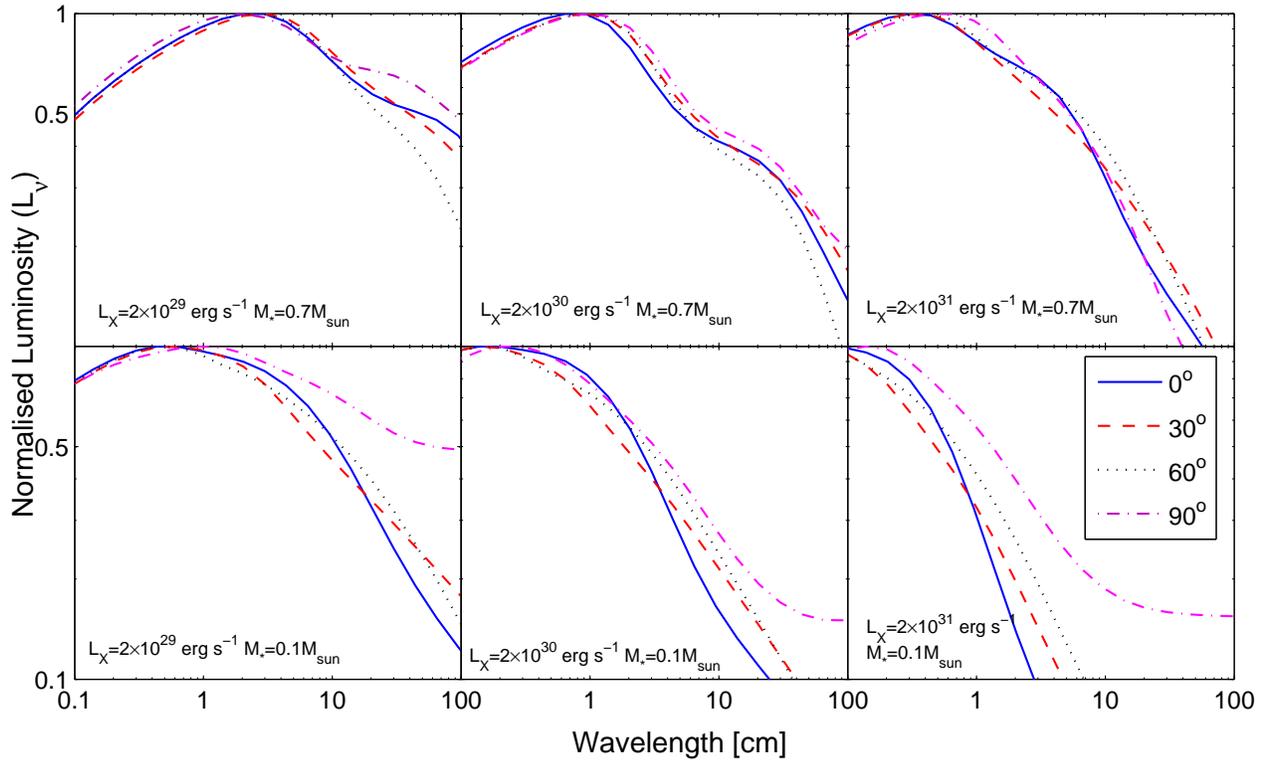}
\caption{Free-Free radio spectrum for the X-ray heated wind at inclination of 0$^o$, 30$^o$, 60$^o$ and 90$^o$. The top/bottom panels show the case of a 0.7/0.1 M$_\odot$ star for X-ray luminosities of $2\times10^{29}$ erg s$^{-1}$ (left), $2\times10^{30}$ erg s$^{-1}$ (middle) and $2\times10^{31}$ erg s$^{-1}$ (right).}\label{fig:xray_spec}
\end{figure*}

In order to compare the expected emission at various wavelengths we compute the spectra in the wavelength range from 0.1 to 100 cm, although we note that disc dust emission is likely to dominate the spectrum at short wavelengths (out to $\sim$ 0.5 cm). We calculate the spectrum at disc inclinations of 0$^o$, 30$^o$, 60$^o$ and 90$^o$. We show these computed spectra for the EUV heated disc in Figure~\ref{fig:euv_spec} (although in this case due to the small dependence on inclination we only plot face-on and edge-on discs) and for the X-ray heated disc in Figure~\ref{fig:xray_spec}. 

The radio spectrum of the disc/wind systems transitions from optically thin at short wavelengths $<10$ cm to optically thick at longer wavelengths $>10$ cm. In general this transition from optically thin to optically thick occurs at shorter wavelength for higher luminosities and around lower mass stars. We can understand these features easily: at higher luminosities the density of the wind increases and thus the opacity also grows; therefore, the transition from optically thin to optically thick shifts to shorter wavelengths. In addition as one moves to lower stellar masses the opacity along a given line of sight increases, this can be easily seen if we write the optical depth in terms of $R_g$ as:
\begin{equation}
\tau_\nu=R_g\int\kappa_\nu{\rm d}\left(r/R_g\right)
\end{equation}
Since $R_g\propto M_*$ and $\kappa_\nu\propto R_g^{-3}$ for the EUV driven wind and as $\kappa_\nu\propto R_g^{-4}$ for the X-ray heated disc, it is clear that the optical depth along a given line of sight increases strongly as the mass decreases for both the X-ray and EUV heated disc.

Furthermore, the two separate emitting regions of the X-ray heated disc results in a bump in the radio spectrum in some cases, arising from the bound atmosphere and wind emitting regions becoming optically thick at different wavelengths. In all cases we find that for similar free-free luminosities, the X-ray spectrum tends to  turn over at shorter wavelengths, and has generally steeper spectral indexes at cm wavelengths. This arises as the dominant emission region in the X-ray case is more compact and is thus more easily attenuated. The impinging EUV flux cannot be directly measured with observations, however a well constrained measurement of the spectral index at cm wavelengths could indirectly constrain the required EUV luminosity, allowing then a comparison between the two photoevaporation models.

\section{A pilot study: GM Aur}
In this section we perform a case study using radio  observations of GM Aur and the formalism developed above. We note that, while GM Aur has a well known inner hole in the dust component, the high accretion rates measured for this object ($\sim 10^{-8}$ M$_\odot$ yr$^{-1}$, Ingelby et al. 2011) imply a substantial amount of gas still present in the inner disc, justifying the application of the primordial disc models developed in the previous section. 

\subsection{Observations of GM Aur}
Observations of GM\,Aur were made during August 2012  with the Arcminute 
Microkelvin Imager Large Array (AMI-LA). The AMI-LA is a synthesis array 
of eight 13\,m antennas sited at the Mullard Radio Astronomy Observatory 
at Lord's Bridge, Cambridge (AMI Consortium: Zwart et~al. 2008). The 
telescope observes in the band $13.5-17.9$\,GHz with eight 0.75\,GHz 
bandwidth channels. In practice channels $1-3$ are generally unused due 
to radio interference from geostationary satellites and data towards 
GM\,Aur for this work were taken from channels $4-8 (14.6-17.9$\,GHz). 
Primary flux calibration was performed using short observations of 3C48 
and 3C286; from other measurements the flux calibration of the array is 
expected to be accurate to better than 5\,per~cent (e.g. AMI Consortium: 
Scaife et~al. 2008). Secondary calibration was carried out using 
interleaved observations of the strong point source J0459+3106.

\begin{figure}
\centering
\includegraphics[width=\columnwidth]{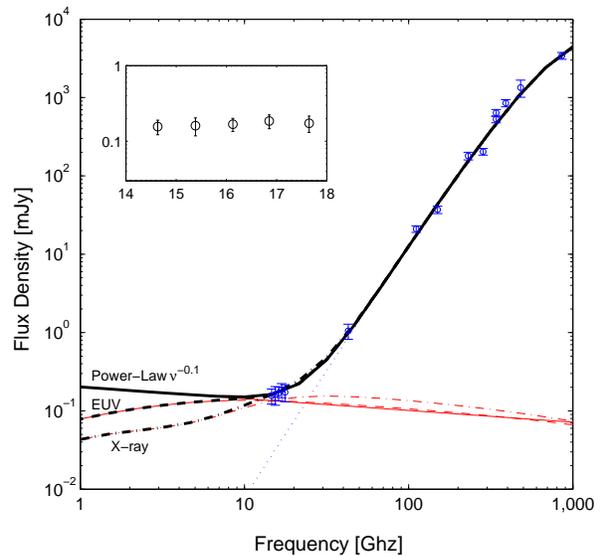}
\caption{Long wavelength observation of GM Aur. The sub-mm and mm observations are taken from Rodmann et al. (2006), Ricci et al. (2010) \& Hughes et al. (2013), whereas the cm observations come from this work; also shown as the insert. The thin dotted line shows the fit to the dust tail, the thin solid, dashed and dot-dashed show  power-law free-free emission, the EUV model and the X-ray model respectfully. The thick lines shows the combined dust+free-free emission spectrum, all which show good agreement with the data.}\label{fig:GM_AUR}
\end{figure}

Calibrated data were imaged using the {\sc aips} data package. With a 
synthesised beam of $39.4''\times25.0''$ towards GMAur, the source is 
unresolved by the AMI-LA. Flux densities were recovered from both the 
combined frequency data, where the source was detected at 
$>10\,\sigma_{\rm rms}$ with $\sigma_{\rm rms}=16\,\mu$Jy\,beam$^{-1}$, 
and individual frequency channels where the source was detected at 
$>5\,\sigma_{\rm rms}$ in each case. No additional sources were detected 
within the AMI-LA primary beam, which has a width at half power of 
$\approx 6$\,arcmin at 16\,GHz. A power-law spectral index was fitted to 
the AMI-LA channel data using the MCMC based Maximum Likelihood 
algorithm {\sc metro} (see e.g. Scaife \& Heald 2012), with a spectral index of -0.76$\pm$0.51 obtained and a radio flux density of 0.13$\pm$0.03 mJy with an estimated $0.04$ mJy contribution from thermal dust emission. We confirm our measurement is of free-free continuum emission rather than an extended dust tail by comparing a pure dust model (single temperature modified black-body), with at two component model (dust+free-free) using the sub-mm and mm data compiled by  Rodmann et al. (2006), Ricci et al. (2010) \& Hughes et al. (2013). This comparison is shown in Figure~\ref{fig:GM_AUR}, where we find the 2 component model (dust+free-free) is very strongly preferred with a Bayes factor 10.50$\pm$0.29 (see Gordon \& Trotta 2007) and derive a dust $\beta$ of 1.26$\pm$0.14, confirming our detection of free-free emission from GM Aur. 

\subsection{Comparison of theory and observations}

\begin{figure}
\includegraphics[width=\columnwidth]{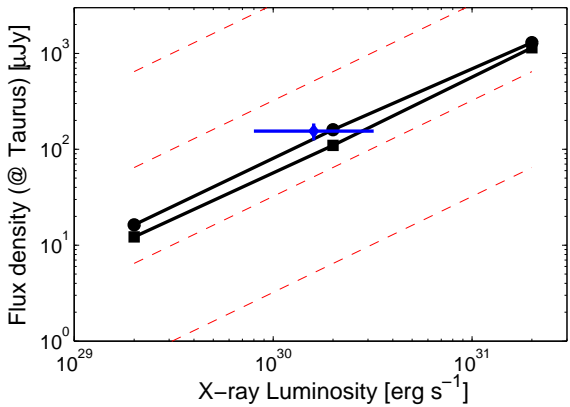}
\caption{The 15GHz free-free flux density at 140pc plotted against X-ray luminosity for a disc observed at a face-on inclination (although inclination makes very little difference at 15~GHz -see Figures~\ref{fig:euv_spec} \& \ref{fig:xray_spec}). The filled squares are for the calculations around a 0.7 M$_\odot$ star and the filled circles are for the calculations around a 0.1 M$_\odot$ star. The dashed lines show the luminosity for the EUV heated wind at EUV luminosities of 0.01, 0.1, 1 \& 10 $L_X$. The blue point shows our GM Aur measurement with the X-ray luminosity taken from Guedel et al. (2010). }\label{fig:luminosity}
\end{figure}

In order to compare our observed value for GM Aur with our theoretical models we show the model luminosity densities at 15~GHz computed for the X-ray models (solid black lines) and the EUV models (dashed) red lines in Figure~\ref{fig:luminosity}. In general 15~GHz roughly represents the transition region from optically thin to optically thick, so the EUV luminosity roughly scales linearly with the free-free flux. Furthermore, since the bound region of the X-ray heated disc dominates the emission, the free-free flux density also scales roughly linearly with X-ray luminosity.  

The blue point in Figure~\ref{fig:luminosity} represents our observed data point for GM Aur and is in good agreement with the expected luminosity from the X-ray model and with the EUV model is one adopts and EUV luminosity of $\sim 10^{41}$ s$^{-1}$, which is  within the expected range for T Tauri stars (Alexander et al. 2005). Thus we cannot use our single data point to distinguish between X-rays and EUV as main heating agents at the surface layers of the GM Aur disc. As well as comparing the luminosity to the models we can also compare the observed spectral index (defined as $-{\rm d}\log L_\nu/{\rm d}\log \nu$), to those obtained from the model calculations. Given GM Aur has a measured X-ray luminosity of $\sim 1.6\times10^{30}$ erg s$^{-1}$ and a stellar mass of $\sim 0.7$ M$_\odot$ (e.g. Guedel et al. 2010), we compare to the 0.7 M$_\odot$ models, in the case of the X-rays we use the $L_X=2\times10^{30}$ erg s$^{-1}$. In the case of the EUV models we show models with luminosities of $10^{40}-10^{42}$ s$^{-1}$, although we note that the observed free-free luminosity implies an EUV luminosity of order $10^{41}$ s$^{-1}$. 
\begin{figure}
\centering
\includegraphics[width=\columnwidth]{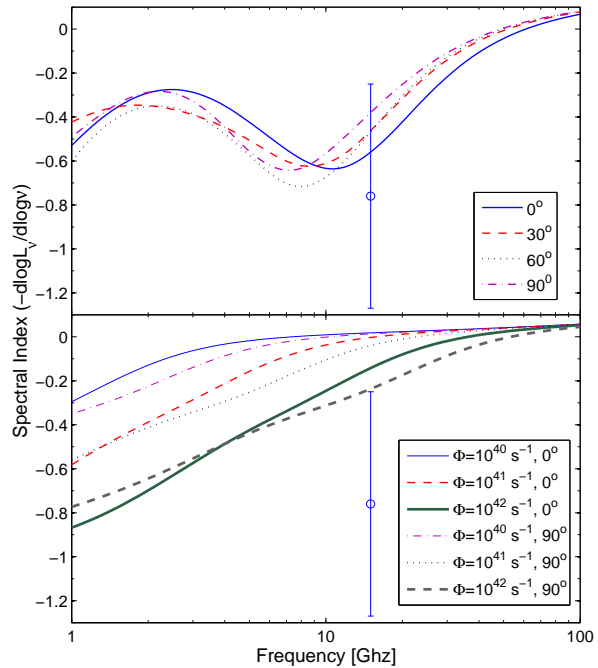}
\caption{Spectral index of the X-ray model (upper panel) and EUV model (bottom panel) compared with the GM Aur measurment (blue-point). The X-ray model is that for a 0.7 M$_\odot$ with and X-ray luminosity of $2\times10^{30}$ erg s$^{-1}$ shown at various inclination. The EUV model is for a 0.7 M$_\odot$ star shown at various inclinations and luminosities.}\label{fig:spec_index}
\end{figure}  
The comparison between the observed spectral index and the models is shown in Figure~\ref{fig:spec_index}, where the X-ray model is in the top panel and the EUV model is in the bottom panel. We find that the X-ray model is a good fit for both the observed free-free flux density and spectral index.  A more accurately determined spectral index at around 10-20 GHz should be able to distinguish between the two models, which predict indices of approximately -0.2 (EUV) and -0.5 (X-ray).  

\subsubsection{Other probes of ionization in GM Aur}
Two further pieces of evidence point to heating of the disc's surface taking place in GM Aur, namely, the observed 6300 \AA~ OI line has a luminosity of $\sim 10^{-5}$ L$_\odot$ (Hartigan et al. 1995), which is believed to have a disc (rather than jet - Guedel et al. 2010) origin is in good agreement with the X-ray model calculations of $1.25\times10^{-5}$ L$_\odot$ from Ercolano \& Owen (2010). Such a high OI luminosity cannot be produced in a EUV heated wind (Font et al. 2004) due to the lack of neutral Oxygen, but may be produced by OH dissociation in the underlying molecular gas discs (Gorti et al. 2011). Furthermore, the 12.8 $\mu$m NeII line has also been detected in GM Aur, with a luminosity of $\sim 7\times10^{-6}$ L$_\odot$ (Najita et al. 2007, Guedel et al. 2010). This also compares favourably with the predicted value from the X-ray models of $5.41\times10^{-6}$ L$_\odot$ from Ercolano \& Owen (2010) and the NeII model which requires a similar EUV flux to the free-free emission to produce the observed luminosity of $\sim 10^{41}$ s$^{-1}$ (Alexander 2008). However, the NeII and free-free emission agreement is not surprising  as both emission features probe the electron density rather than the total density. {\y Neither the OI or NeII line show evidence for a blue-shift in GM Aur's disc. However, the lack of blue-shift does not necessarily indicate the absence of a photoevaporative wind in GM Aur. The dust disc of GM
Aur has a large inner hole, allowing the 
red-shifted side of the wind to become visible to the observer, hence making the line profile symmetric,  or allow
the bound component of the heated disc to dominant the
emission (Ercolano \& Owen, 2010).}  

\section{Discussion}\label{sec:discuss}
In the previous sections we have shown that free-free emission from the hot heated atmospheres of protoplanetary discs, both in the bound inner regions and photoevaporative wind at larger distances can provide a useful probe of the high energy spectrum impinging upon the discs surface. Combing the continuum radio emission measurement with further probes of the photoevaporative wind such as gas line emission (e.g. Font et al. 2004; Alexander 2008; Pascucci et al. 2009; Ercolano \& Owen 2010, Pascucci et al. 2011, Szul{\'a}gyi et al. 2012, Espaillat et al. 2013) should provide useful constraints on the properties of the photoevaporative flow in individual systems. 

\begin{figure*}
\centering
\includegraphics[width=\textwidth]{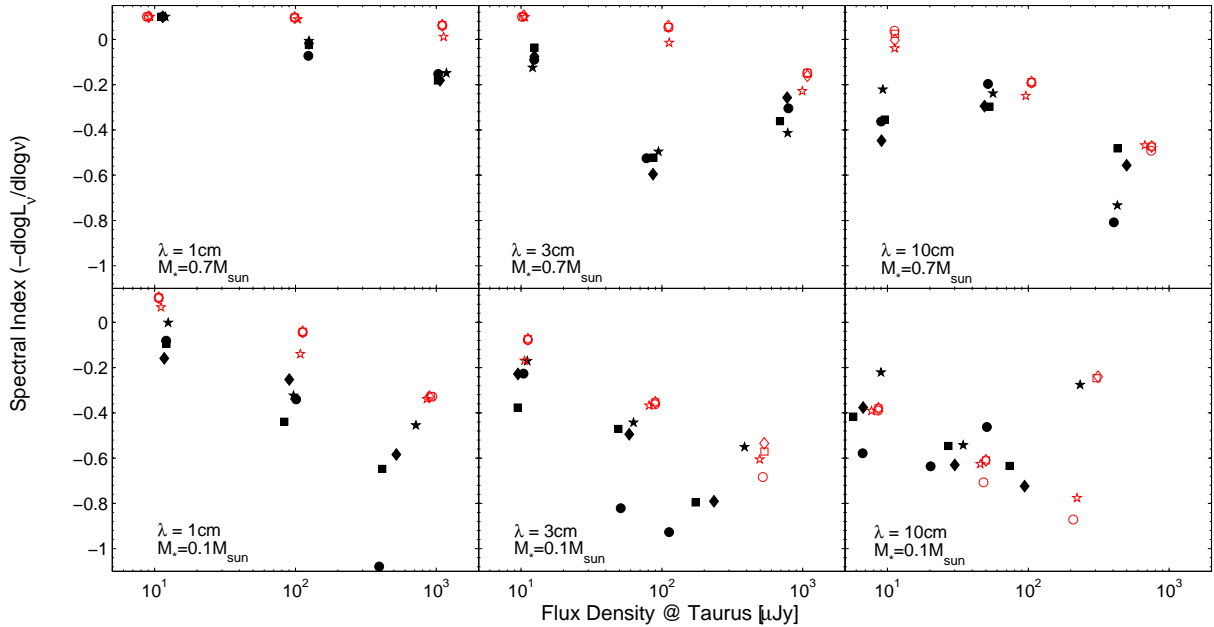}
\caption{Spectral index of the calculations presented in Section 3, plotted as a function of free-free flux density. The open symbols represent the EUV model and the filled symbols the X-ray model.  Circular, Square, Diamond and Stars show the results when observed at different disc inclinations of 0, 30, 60 \& 90 degrees respectfully. }\label{fig:si_flux}
\end{figure*}

In the previous section we have shown that a pilot measurement for GM Aur and compared to our models. While in this case this exercise did not yield a definitive answer to the question of what may be the dominant (EUV or X-ray) heating/ionisation agent for this disc, it has shown how more accurate spectral index measurement may in the future succeed in this task. {\y Exploring the combination of flux density and spectral index measurements further we plot the expected free-free flux densities and spectral indices of all calculated models in Figure~\ref{fig:si_flux} at three representative wavelengths of 1, 3 \& 10 cm. Figure~\ref{fig:si_flux} clearly shows that a well determined spectral index (to within $\sim 0.2$ dex) would be able to distinguish between the two models at the shorter radio wavelengths (i.e. 1 \& 3 cm), this is where the EUV model is still essentially optically thin but the X-ray model is beginning to become optically thick. Coupled with X-ray luminosity measurements, which should constrain the expected value of the free-free flux density (in the X-ray model) and multi-wavelength measurements the dominant driving source of the photoevaporative wind could be determined in individual sources with current instrumentation. }
In the next section we present a framework for testing the dominant driver of photoevaporation, based on the previous considerations.

\subsection{A framework for testing photoevaporation}
Given the similarity between the predicted free-free luminosities and spectral indices from the EUV and X-ray photoevaporation models it may seem unlikely that free-free emission could prove a useful probe alone or even when combined with the 12.8$\mu$m NeII line. However, since the free-free luminosities are similar, but the mass-loss rates and mass-loss rate scalings vary greatly between the two models, armed with additional, widely accessible, measurements of accretion  rates ($\dot{M}_*$) one can construct a powerful diagnostic for testing the photoevaprative disc destruction scenarios and address the question of whether X-rays or EUV radiation is the driving mechanism. 

In particular a EUV driven wind results in a much lower mass-loss rate and a weak $\Phi^{1/2}$ scaling with luminosity, compared to the higher mass-loss rates and linear scaling with luminosity for the X-ray driven case. One can then adopt a zeroth order approximation in the context of photoevaporation (Clarke et al. 2001), which states that the dispersal proceeds when the accretion rates and mass-loss rates are equal.  In this case at the point of disc dispersal one finds that the free-free luminosity scales as $\dot{M}_*^2$ in the case of an EUV driven wind and as $\dot{M}_*$ in the case of and X-ray driven wind. 

Using these two scalings and the knowledge that a disc disperses due to photoevaporation once the accretion rates drops below the mass-loss rate, a simple experiment can be constructed as follows. One measures the free-free flux density for a large sample of stars at different accretion rates and plots the free-free flux density as a function of accretion rate, which should show a lack of objects with large free-free luminosities and low accretion rates, due to the fact that stars with high free-free luminosities (i.e. high EUV/X-ray flux) will have already dispersed their discs due to their high mass-loss rates. 

{\y In order to probe this hypothesis further, we use a set of evolving disc models undergoing viscous evolution and photoevaporation. We do this by solving the viscous diffusion equation with mass-loss (Clarke et al. 2001) for a population of discs:
\begin{equation}
\frac{\partial\Sigma}{\partial t}=\frac{3}{R}\frac{\partial}{\partial R}\left[R^{1/2}\frac{\partial}{\partial R}\left(\nu R^{1/2}\Sigma\right)\right]-\dot{\Sigma_w}\label{eqn:viscous_evolve}
\end{equation}
 where $\Sigma$ is the gas surface density, $\nu$ is the viscosity and $\dot{\Sigma}_w$ represents the mass-loss due to photoevaporation. We use the standard viscosity scaling $\nu\propto R$ (Alexander et al. 2006; Alexander \& Armitage 2007;2009; Owen et al. 2010;2011b) and adopt a zero time Lynden-Bell \& Pringle similarity solution as our initial surface density profile (Lynden-Bell \& Pringle 1974). 
 
 Equation~\ref{eqn:viscous_evolve} is integrated forward using the scheme described in detail in Owen et al. (2011b) for a disc around a 0.7 M$_\odot$ star. For the X-ray driven photoevaporation we use the same initial conditions as Owen et al. (2011b) of an initial disc mass of 0.07 M$_\odot$, an initial disc scale radius of 18 AU and a viscous $\alpha$ of $2\times10^{-3}$, which were chosen to match the observed disc lifetimes. Where we use the X-ray photoevaporation profile presented in Owen et al. (2012). For the EUV model we use the same initial conditions as Alexander \& Armitage (2009), again which were originally chosen to match the observed disc lifetimes. This were an initial disc mass of $\log_{10}(M_d(t=0)/M_*)=-1.5$, an initial disc scale radius of 10 AU and a viscous $\alpha$ of 0.01, where we used the EUV photoevaporation profile found by Font et al. (2004) and presented in Alexander \& Armitage (2007). We emphasise that given we are interested in relative time-scales the results are somewhat insensitive to the actual choice of initial conditions (Alexander et al. 2006; Owen et al. 2011b). For the X-ray models we span the observed X-ray luminosity function of solar-like stars (Guedel et al. 2007), and cover the same range in free-free luminosity for the EUV model (roughly $\Phi_*=10^{40}-10^{42}$ s$^{-1}$).

\begin{figure}
\includegraphics[width=0.9\columnwidth]{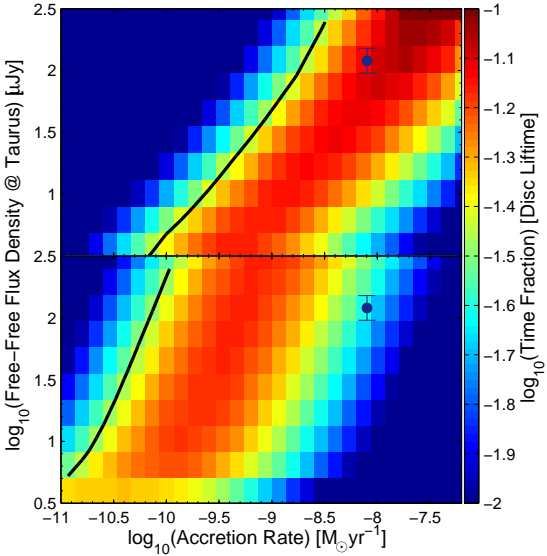}
\caption{Map of time-spent in a certain region of parameter space by viscously evolving disc models with photoevaporation, plotted normalised to the disc's lifetime. The top panel is for the X-ray driven winds, and the bottom panel for the EUV driven winds. The solid lines represent the region of parameter that will be probed upto by a sample of a few hundred  stars. The point in our GM Aur detection.}\label{fig:region}
\end{figure}
In Figure~\ref{fig:region} we show the results of these calculations, plotted as the time-spent by a model in a certain region of the parameter space in units of disc lifetime. The X-ray model is shown in the top panel and the EUV in the bottom panel, with our GM Aur data point also shown. We cannot produce an expected probability map as the EUV luminosity function is unknown; however, given a sample of roughly 300 young disc-bearing systems (roughly the number with FIR/submm measurements needed to determine a free-free excess), then one would expect to probe areas to within a few percent of an individual disc model's life. This expected line is shown as the black line in Figure~\ref{fig:region} .  Where the slopes of this line roughly correspond to the free-free luminosity scaling predicted by the EUV and X-ray models of $L_\nu^{\rm EUV}\propto\dot{M}_*^2$ and $L_\nu^{X}\propto\dot{M}_*$, respectively. }

Transition discs, which should represent the last {\y rapid stage ($\sim 10\%$ of disc's lifetime)} of disc evolution, are particularly useful for such a survey. However, we also note that one must be careful when selecting `transition' discs, since many objects that may be identified as `transition' disc due to the fact that they have an inner hole/gap, but may not actually be discs caught in the act of transitioning between a disc-bearing and disc-less state (Owen \& Clarke 2012). In particular, GM AUR itself may fall into this case, as a disc being identified as a `transition' disc (e.g. Calvet et al. 2002; Andrews et al. 2011), but where the high inferred disc mass and accretion rate raises imply its hole may have been dynamically carved by an embedded giant planet (e.g. Calvet et al. 2005, Rice et al. 2006, Andrews et al. 2011) rather than due to photoevaporation, although problems still exist with this latter interpretation (Clarke \& Owen 2013).

Nevertheless a carefully selected sample of `transition discs' along with a sample of primordial discs should be able to provide a strong initial test. Combining measurements of free-free luminosity and spectral index at frequencies of $1- 10$ GHz, should provide useful complimentary evidence to the line emission to further test and constrain the photoevaporation scenario.

\section{Summary}
We have considered the suggestion of Pascucci et al. (2012) to use free-free radio continuum emission to explore the irradiation of protoplanetary discs by EUV and X-ray irradiation. We use previously computed hydrodynamic simulations of EUV \& X-ray heated discs to calculate the expected properties of photoevaporating discs when observed at radio frequencies. In general, we find that the free-free luminosity scales linearly with the EUV or X-ray luminosity of the central star and should be detectable at cm wavelengths above the dust tail of the protoplanetary disc. In particular, we use the T Tauri star GM Aur as a test case and detect free-free emission at the level predicted by the models, although we are unable to use the current data to rule out either the EUV or X-ray photoevaporation model. Our main findings are highlighted below:
\begin{enumerate}
\item Free-free emission from the ionized disc's atmosphere dominates at cm wavelengths. Within the EUV photoevaporation model, it is the wind that dominated the emission; however, in the X-ray model it is the hot bound atmosphere close to the star ($<$ 1 AU) that dominates the free-free emission.

\item Both the X-ray and EUV heating models predict the free-free flux density should scale roughly linearly with ionizing luminosity, and should be detectable with current radio telescopes.  

\item We have detected radio emission from GM Aur consistent with free-free emission from the ionized surface layers of the inner gas disc.

\item A radio survey of local primordial discs and carefully selected transition discs, combined with measurements of the mass-accretion maybe able to distinguish between the EUV and X-ray photoevaporation scenarios.  

\end{enumerate}

\section*{Acknowledgements}
We thank the anonymous referee for suggestions which helped improve the paper. We thank Cathie Clarke, Jermery Drake, Ilaria Pascucci, Richard Alexander and Yanqin Wu for interesting discussions. The calculations were performed
on the Sunnyvale cluster at CITA which is funded by the
Canada Foundation for Innovation. We thank the staff of the Lord’s Bridge Observatory for their invaluable assistance in the commissioning and operation of the Arcminute Microkelvin
Imager. AMI-LA is supported by Cambridge University and the STFC.


\begin{thebibliography}{99}

\bibitem[\protect\citeauthoryear{Adams et al.}{2004}]{2004ApJ...611..360A} 
Adams F.~C., Hollenbach D., Laughlin G., Gorti U., 2004, ApJ, 611, 360 

\bibitem[\protect\citeauthoryear{Alexander, Clarke, 
\& Pringle}{2005}]{2005MNRAS.358..283A} Alexander R.~D., Clarke C.~J., Pringle J.~E., 2005, MNRAS, 358, 283

\bibitem[\protect\citeauthoryear{Alexander, Clarke, 
\& Pringle}{2006}]{2006MNRAS.369..216A} Alexander R.~D., Clarke C.~J., Pringle J.~E., 2006, MNRAS, 369, 216
\bibitem[\protect\citeauthoryear{Alexander 
\& Armitage}{2007}]{2007MNRAS.375..500A} Alexander R.~D., Armitage P.~J., 2007, MNRAS, 375, 500

\bibitem[\protect\citeauthoryear{Alexander}{2008}]{2008MNRAS.391L..64A} 
Alexander R.~D., 2008, MNRAS, 391, L64 

\bibitem[\protect\citeauthoryear{Alexander 
\& Armitage}{2009}]{2009ApJ...704..989A} Alexander R.~D., Armitage P.~J., 2009, ApJ, 704, 989


\bibitem[\protect\citeauthoryear{Andrews et 
al.}{2011}]{2011ApJ...732...42A} Andrews S.~M., Wilner D.~J., Espaillat C., 
Hughes A.~M., Dullemond C.~P., McClure M.~K., Qi C., Brown J.~M., 2011, 
ApJ, 732, 42

\bibitem[\protect\citeauthoryear{Armitage 
\& Hansen}{1999}]{1999Natur.402..633A} Armitage P.~J., Hansen B.~M.~S., 1999, Natur, 402, 633 


\bibitem[Birnstiel et al.(2012)]{2012A&A...544A..79B} Birnstiel, T., Andrews, S.~M., \& Ercolano, B.\ 2012, \aap, 544, A79 



\bibitem[\protect\citeauthoryear{Brown et al.}{2009}]{2009ApJ...704..496B} 
Brown J.~M., Blake G.~A., Qi C., Dullemond C.~P., Wilner D.~J., Williams 
J.~P., 2009, ApJ, 704, 496 



\bibitem[\protect\citeauthoryear{Calvet et al.}{2002}]{2002ApJ...568.1008C} 
Calvet N., D'Alessio P., Hartmann L., Wilner D., Walsh A., Sitko M., 2002, 
ApJ, 568, 1008

\bibitem[\protect\citeauthoryear{Calvet et al.}{2005}]{2005ApJ...630L.185C} 
Calvet N., et al., 2005, ApJ, 630, L185 


\bibitem[\protect\citeauthoryear{Clarke, Gendrin, 
\& Sotomayor}{2001}]{2001MNRAS.328..485C} Clarke C.~J., Gendrin A., Sotomayor M., 2001, MNRAS, 328, 485
\bibitem[\protect\citeauthoryear{Clarke \& Owen}{2013}]{inset} Clarke C.~J., Owen J.~E., 2013 MNRAS {\it in press}

\bibitem[Dullemond 
\& Dominik(2005)]{2005A&A...434..971D} Dullemond, C.~P., \& Dominik, C.\ 2005, \aap, 434, 971 

\bibitem[\protect\citeauthoryear{Ercolano et 
al.}{2003}]{2003MNRAS.340.1136E} Ercolano B., Barlow M.~J., Storey P.~J., 
Liu X.-W., 2003, MNRAS, 340, 1136 

\bibitem[\protect\citeauthoryear{Ercolano, Barlow, 
\& Storey}{2005}]{2005MNRAS.362.1038E} Ercolano B., Barlow M.~J., Storey P.~J., 2005, MNRAS, 362, 1038 

\bibitem[\protect\citeauthoryear{Ercolano et 
al.}{2008}]{2008ApJ...688..398E} Ercolano B., Drake J.~J., Raymond J.~C., 
Clarke C.~C., 2008, ApJ, 688, 398

\bibitem[\protect\citeauthoryear{Ercolano 
\& Owen}{2010}]{2010MNRAS.406.1553E} Ercolano B., Owen J.~E., 2010, MNRAS, 406, 1553 

\bibitem[Ercolano et al.(2011)]{2011MNRAS.416..439E} Ercolano, B., Bastian, 
N., Spezzi, L., \& Owen, J.\ 2011, \mnras, 416, 439 

\bibitem[\protect\citeauthoryear{Espaillat et 
al.}{2010}]{2010ApJ...717..441E} Espaillat C., et al., 2010, ApJ, 717, 441 

\bibitem[\protect\citeauthoryear{Espaillat et 
al.}{2013}]{2013ApJ...762...62E} Espaillat C., et al., 2013, ApJ, 762, 62 

\bibitem[\protect\citeauthoryear{Ercolano, Clarke, 
\& Hall}{2011}]{2011MNRAS.410..671E} Ercolano B., Clarke C.~J., Hall A.~C., 2011, MNRAS, 410, 671
\bibitem[\protect\citeauthoryear{Feigelson et 
al.}{1994}]{1994ApJ...432..373F} Feigelson E.~D., Welty A.~D., Imhoff C., 
Hall J.~C., Etzel P.~B., Phillips R.~B., Lonsdale C.~J., 1994, ApJ, 432, 
373 
\bibitem[\protect\citeauthoryear{Font et al.}{2004}]{2004ApJ...607..890F} 
Font A.~S., McCarthy I.~G., Johnstone D., Ballantyne D.~R., 2004, ApJ, 607, 
890
\bibitem[\protect\citeauthoryear{Forbrich et 
al.}{2007}]{2007A&A...464.1003F} Forbrich J., et al., 2007, A\&A, 464, 1003 

\bibitem[\protect\citeauthoryear{Forbrich, Osten, 
\& Wolk}{2011}]{2011ApJ...736...25F} Forbrich J., Osten R.~A., Wolk S.~J., 2011, ApJ, 736, 25

\bibitem[\protect\citeauthoryear{Forbrich 
\& Wolk}{2013}]{2013A&A...551A..56F} Forbrich J., Wolk S.~J., 2013, A\&A, 551, A56

\bibitem[\protect\citeauthoryear{Gagn{\'e}, Skinner, 
\& Daniel}{2004}]{2004ApJ...613..393G} Gagn{\'e} M., Skinner S.~L., Daniel K.~J., 2004, ApJ, 613, 393 
\bibitem[\protect\citeauthoryear{Gorti et al.}{2011}]{2011ApJ...735...90G} 
Gorti U., Hollenbach D., Najita J., Pascucci I., 2011, ApJ, 735, 90

\bibitem[\protect\citeauthoryear{Gordon 
\& Trotta}{2007}]{2007MNRAS.382.1859G} Gordon C., Trotta R., 2007, MNRAS, 382, 1859 


\bibitem[\protect\citeauthoryear{Gorti 
\& Hollenbach}{2009}]{2009ApJ...690.1539G} Gorti U., Hollenbach D., 2009, ApJ, 690, 1539

\bibitem[\protect\citeauthoryear{Gorti, Dullemond, 
\& Hollenbach}{2009}]{2009ApJ...705.1237G} Gorti U., Dullemond C.~P., Hollenbach D., 2009, ApJ, 705, 1237

\bibitem[\protect\citeauthoryear{Guedel 
\& Benz}{1993}]{1993ApJ...405L..63G} Guedel M., Benz A.~O., 1993, ApJ, 405, L63
\bibitem[\protect\citeauthoryear{G{\"u}del et 
al.}{2007}]{2007A&A...468..353G} G{\"u}del M., et al., 2007, A\&A, 468, 353
\bibitem[\protect\citeauthoryear{G{\"u}del et 
al.}{2010}]{2010A&A...519A.113G} G{\"u}del M., et al., 2010, A\&A, 519, A113 
\bibitem[\protect\citeauthoryear{Guenther et 
al.}{2000}]{2000A&A...357..206G} Guenther E.~W., et al., 2000, A\&A, 357, 206

\bibitem[\protect\citeauthoryear{Haisch, Lada, 
\& Lada}{2001}]{2001ApJ...553L.153H} Haisch K.~E., Jr., Lada E.~A., Lada C.~J., 2001, ApJ, 553, L153 

\bibitem[\protect\citeauthoryear{Hartigan, Edwards, 
\& Ghandour}{1995}]{1995ApJ...452..736H} Hartigan P., Edwards S., Ghandour L., 1995, ApJ, 452, 736 

\bibitem[\protect\citeauthoryear{Hern{\'a}ndez et 
al.}{2007}]{2007ApJ...671.1784H} Hern{\'a}ndez J., et al., 2007, ApJ, 671, 
1784 
\bibitem[\protect\citeauthoryear{Hughes et al.}{2013}]{2013AJ....145..115H} 
Hughes A.~M., Hull C.~L.~H., Wilner D.~J., Plambeck R.~L., 2013, AJ, 145, 
115 
\bibitem[\protect\citeauthoryear{Hollenbach et 
al.}{1994}]{1994ApJ...428..654H} Hollenbach D., Johnstone D., Lizano S., 
Shu F., 1994, ApJ, 428, 654 


\bibitem[\protect\citeauthoryear{Johnstone, Hollenbach, 
\& Bally}{1998}]{1998ApJ...499..758J} Johnstone D., Hollenbach D., Bally J., 1998, ApJ, 499, 758 

\bibitem[\protect\citeauthoryear{Kenyon 
\& Hartmann}{1995}]{1995ApJS..101..117K} Kenyon S.~J., Hartmann L., 1995, ApJS, 101, 117 

\bibitem[\protect\citeauthoryear{Kim et al.}{2009}]{2009ApJ...700.1017K} 
Kim K.~H., et al., 2009, ApJ, 700, 1017



\bibitem[\protect\citeauthoryear{Koepferl et 
al.}{2013}]{2013MNRAS.428.3327K} Koepferl C.~M., Ercolano B., Dale J., 
Teixeira P.~S., Ratzka T., Spezzi L., 2013, MNRAS, 428, 3327


\bibitem[\protect\citeauthoryear{Luhman et al.}{2010}]{2010ApJS..186..111L} 
Luhman K.~L., Allen P.~R., Espaillat C., Hartmann L., Calvet N., 2010, 
ApJS, 186, 111 

\bibitem[\protect\citeauthoryear{Lynden-Bell 
\& Pringle}{1974}]{1974MNRAS.168..603L} Lynden-Bell D., Pringle J.~E., 1974, MNRAS, 168, 603

\bibitem[\protect\citeauthoryear{Mamajek}{2009}]{2009AIPC.1158....3M} 
Mamajek E.~E., 2009, AIPC, 1158, 3 

\bibitem[\protect\citeauthoryear{Mer{\'{\i}}n et 
al.}{2010}]{2010ApJ...718.1200M} Mer{\'{\i}}n B., et al., 2010, ApJ, 718, 
1200 

\bibitem[\protect\citeauthoryear{M{\"u}cke et 
al.}{2002}]{2002ApJ...571..366M} M{\"u}cke A., Koribalski B.~S., Moffat 
A.~F.~J., Corcoran M.~F., Stevens I.~R., 2002, ApJ, 571, 366 


\bibitem[\protect\citeauthoryear{Osten 
\& Wolk}{2009}]{2009ApJ...691.1128O} Osten R.~A., Wolk S.~J., 2009, ApJ, 691, 1128

\bibitem[\protect\citeauthoryear{Owen et al.}{2010}]{2010MNRAS.401.1415O} 
Owen J.~E., Ercolano B., Clarke C.~J., Alexander R.~D., 2010, MNRAS, 401, 
1415

\bibitem[\protect\citeauthoryear{Owen, Ercolano, 
\& Clarke}{2011b}]{2011MNRAS.412...13O} Owen J.~E., Ercolano B., Clarke C.~J., 2011, MNRAS, 412, 13

\bibitem[\protect\citeauthoryear{Owen, Ercolano, 
\& Clarke}{2011a}]{2011MNRAS.411.1104O} Owen J.~E., Ercolano B., Clarke C.~J., 2011, MNRAS, 411, 1104

\bibitem[Owen et al.(2012)]{2012MNRAS.422.1880O} Owen, J.~E., Clarke, 
C.~J., \& Ercolano, B.\ 2012, \mnras, 422, 1880 

\bibitem[\protect\citeauthoryear{Owen 
\& Clarke}{2012}]{2012MNRAS.426L..96O} Owen J.~E., Clarke C.~J., 2012, MNRAS, 426, L96 


\bibitem[\protect\citeauthoryear{Pascucci 
\& Sterzik}{2009}]{2009ApJ...702..724P} Pascucci I., Sterzik M., 2009, ApJ, 702, 724 


\bibitem[\protect\citeauthoryear{Pascucci et 
al.}{2011}]{2011ApJ...736...13P} Pascucci I., et al., 2011, ApJ, 736, 13 

\bibitem[\protect\citeauthoryear{Pascucci, Gorti, 
\& Hollenbach}{2012}]{2012ApJ...751L..42P} Pascucci I., Gorti U., Hollenbach D., 2012, ApJ, 751, L42

\bibitem[\protect\citeauthoryear{Preibisch et 
al.}{2005}]{2005ApJS..160..401P} Preibisch T., et al., 2005, ApJS, 160, 401

\bibitem[\protect\citeauthoryear{Ricci et 
al.}{2010}]{2010A&A...512A..15R} Ricci L., Testi L., Natta A., Neri R., Cabrit S., Herczeg G.~J., 2010, A\&A, 512, A15

\bibitem[\protect\citeauthoryear{Rice et al.}{2006}]{2006MNRAS.373.1619R} 
Rice W.~K.~M., Armitage P.~J., Wood K., Lodato G., 2006, MNRAS, 373, 1619

\bibitem[\protect\citeauthoryear{Richling 
\& Yorke}{2000}]{2000ApJ...539..258R} Richling S., Yorke H.~W., 2000, ApJ, 539, 258

\bibitem[\protect\citeauthoryear{Rodmann et 
al.}{2006}]{2006A&A...446..211R} Rodmann J., Henning T., Chandler C.~J., Mundy L.~G., Wilner D.~J., 2006, A\&A, 446, 211

\bibitem[\protect\citeauthoryear{Rosotti et 
al.}{2013}]{2013MNRAS.430.1392R} Rosotti G.~P., Ercolano B., Owen J.~E., 
Armitage P.~J., 2013, MNRAS, 430, 1392

\bibitem[\protect\citeauthoryear{Scaife et al.}{2008}]{2008MNRAS.385..809S}
Scaife A.~M.~M., et al., 2008, MNRAS, 385, 809

\bibitem[\protect\citeauthoryear{Scaife et 
al.}{2011}]{2011MNRAS.415..893A} Scaife, et al., 2011, MNRAS, 415, 
893 
\bibitem[\protect\citeauthoryear{Scaife
\& Heald}{2012}]{2012MNRAS.423L..30S} Scaife A.~M.~M., Heald G.~H., 2012,
MNRAS, 423, L30


\bibitem[\protect\citeauthoryear{Skrutskie et 
al.}{1990}]{1990AJ.....99.1187S} Skrutskie M.~F., Dutkevitch D., Strom 
S.~E., Edwards S., Strom K.~M., Shure M.~A., 1990, AJ, 99, 1187 

\bibitem[\protect\citeauthoryear{Strom et al.}{1989}]{1989AJ.....97.1451S} 
Strom K.~M., Strom S.~E., Edwards S., Cabrit S., Skrutskie M.~F., 1989, AJ, 
97, 1451

\bibitem[\protect\citeauthoryear{Szul{\'a}gyi et 
al.}{2012}]{2012ApJ...759...47S} Szul{\'a}gyi J., Pascucci I., 
{\'A}brah{\'a}m P., Apai D., Bouwman J., Mo{\'o}r A., 2012, ApJ, 759, 47



\bibitem[\protect\citeauthoryear{Zwart et al.}{2008}]{2008MNRAS.391.1545Z}
Zwart J.~T.~L., et al., 2008, MNRAS, 391, 1545


\end{thebibliography}
\end{document}